\RequirePackage[l2tabu, orthodox]{nag}
\documentclass[%
twocolumn,
superscriptaddress,
showpacs,
 amsmath,amssymb,
 aps,
pre,
]{revtex4-1}
\usepackage{tikz}
\usepackage{graphicx}
\usepackage{amsfonts}
\usepackage{amsmath}
\usepackage{amssymb}
\usepackage{dcolumn}
\usepackage{bm}
\usepackage{color}
\usepackage[caption=false]{subfig}

\usetikzlibrary{shapes.geometric,calc}
\begin{document}	
	
\preprint{APS/123-QED}

\title{Electromagnetic-Magnetoelectric Duality for Waveguides}
\author{Nafiseh Sang-Nourpour}
\affiliation{Photonics Group, Research Institute for Applied Physics and Astronomy, University of Tabriz, Tabriz 51665-163, Iran}
\affiliation{Institute for Quantum Science and Technology, University of Calgary, Alberta T2N 1N4, Canada}
\affiliation{Photonics Group, Aras International Campus of University of Tabriz, Tabriz 51666-16471, Iran}
\author{Benjamin R. Lavoie}
\affiliation{Department of Electrical and Computer Engineering, Schulich School of Engineering, University of Calgary, Alberta T2N 1N4, Canada}
\affiliation{Institute for Quantum Science and Technology, University of Calgary, Alberta T2N 1N4, Canada}
\author{R. Kheradmand}
\affiliation{Photonics Group, Research Institute for Applied Physics and Astronomy, University of Tabriz, Tabriz 51665-163, Iran}
\author{M. Rezaei}
\affiliation{Department of Theoretical Physics, Faculty of Physics, University of Tabriz, Tabriz 51664, Iran}
\affiliation{Department of Physics, Applied Science Faculty, Simon Fraser University, Burnaby, Canada}
\author{Barry C. Sanders}
\affiliation{Institute for Quantum Science and Technology, University of Calgary, Alberta T2N 1N4, Canada}
\affiliation{Hefei National Laboratory for Physics at Microscale and Department of Modern Physics, University of Science and Technology of China, Hefei, Anhui, China}
\affiliation{Shanghai Branch, CAS Center for Excellence and Synergetic Innovation Center in Quantum Information and Quantum Physics, University of Science and Technology of China, Shanghai 201315, China}
\affiliation{Program in Quantum Information Science, Canadian Institute for Advanced Research, Toronto, Ontario M5G 1Z8, Canada}

\begin{abstract}

We develop a theory for waveguides that respects the duality of electromagnetism,
namely the symmetry of the equations arising through inclusion of magnetic monopoles in addition to including electrons (electric monopoles).
The term magnetoelectric potential is sometimes used to signify the magnetic-monopole induced dual to the usual electromagnetic potential.
To this end, we introduce a general theory for describing modes and characteristics of waveguides based on mixed-monopole materials, with both electric and magnetic responses. Our theory accommodates exotic media such as double-negative, near-zero and zero-index materials, and we demonstrate that our general theory exhibits the electromagnetic duality that would arise if we were to incorporate magnetic monopoles into the media. We consider linear, homogeneous, isotropic waveguide materials with slab and cylindrical geometries. To ensure manifest electromagnetic duality, we construct generic electromagnetic susceptibilities that are dual in both electric charges and magnetic monopoles using a generalized Drude-Lorentz model that manifests this duality.
Our model reduces to standard cases in appropriate limits. We consider metamaterials and metamaterial waveguides, as well as metal guides, as examples of waveguides constructed of mixed-monopole materials, to show the generality of our waveguide theory. In particular, we show that, in slab and cylindrical waveguides, exchanging electric and magnetic material properties leads to the exchange of transverse magnetic and transverse electric modes and dispersion equations, which suggests a good test of the potential duality of waveguides. This theory has the capability to predict waveguide behavior under an exchange of electromagnetic parameters from the dual waveguide.  

\end{abstract}
\maketitle

\section{\label{Introduction}Introduction}

Electromagnetic waveguides are physical devices that confine and control the energy of propagating waves, and are able to efficiently direct electromagnetic (EM) radiation~\cite{Yeh08}. The applications of these structures, such as in radars and optical fibers, provide motivation for further advances in this field.
In recent years, metamaterials~\cite{Shalaev2007} have enriched the capabilities of waveguides, for example by tailoring properties such as EM susceptibilities as well as by generating new kinds of modes~\cite{Shadrivov2003, Lavoie2012}. Metamaterials are designed to have both negative and positive EM susceptibilities within a certain frequency window,
and can even have both susceptibilities negative in some window,
which is the so-called double-negative medium~\cite{Engheta2006, Ramakrishna2005, Shalaev2007}.

Although breakthroughs are being made with custom-designed susceptibilities using metamaterials,
which are synthesized as some specific geometric arrangement of materials to realize electric and magnetic properties that might not match naturally available materials
~\cite{Shalaev2007},
here we treat metamaterials as materials in their own right at both macroscopic and microscopic levels.
Our material description of metamaterials is natural at the macroscopic level where the purpose of the metamaterial is to behave as an effective material on a sufficiently large scale.
On the microscopic level,
the material description is advantageous,
as we shall see,
because we can employ microscopic models such as the Drude-Lorentz model for calculating electric and magnetic responses of the medium~\cite{Jackson1999}.

In order to describe exotic metamaterials, for example with double-negative index properties,
as materials per se,
our description of the material must manifest full electromagnetic duality.
This electromagnetic duality arises by incorporating magnetic monopoles into the electromagnetic equations.
One way to admit monopoles is to construct two potentials,
namely the usual electromagnetic potential arising from Maxwell's equations,
and a second potential,
introduced by Cabibbo and Ferrari~\cite{CF62,MR75}
and sometimes referred to as the magnetoelectric potential~\cite{Mou01,Fry89}.
Whereas a metamaterial can be synthesized from regular materials to yield exotic properties such as double-negative indices,
introducing magnetic monopoles into the theory enables a purely microscopic description of these media as materials rather than metamaterials.
We find the fully dual description of electromagnetism conducive to building both the formalism and the intuition for a fully general theory of waveguides with utility for designing metamaterial waveguides as well as material waveguides.

Using this convenient formal description of materials in the framework of fully dualistic electromagnetic theory,
we can generalize waveguide theories to accommodate this fully general setting.
Although impressive efforts have generalized waveguide descriptions for various
geometries and constituent materials~\cite{Yeh08, Hyuck2012, Halterman2007, Lavoie2012},
a fully comprehensive theory of waveguides for all materials and metamaterials is lacking.
Our aim is to construct a general waveguide theory incorporating full duality and showing how this theory can convert descriptions of ordinary waveguides into exotic ones simply using the principle of duality.
We focus on duality for waveguides comprising linear, homogeneous, isotropic materials
(including exotic materials with magnetic monopoles) and do not deal with varying the geometry
at this time;
a general theory incorporating full duality as well as general geometries is challenging but could be quite interesting given the importance and challenges of conformal transformation optics~\cite{XC15}.

Our generalized waveguide theory features the full duality inherent in Maxwell's equations and gives a quantitative description of waveguide phenomena, both macroscopically and microscopically. The relevant background of the theory is presented in Sec.~\textrm{II}. We detail our approach in Sec.~\textrm{III}. The mathematical results and examples are presented in Sec.~\textrm{IV}. We conclude with a discussion of our theory in Sec.~\textrm{V}.

\section{\label{Background}Background}

This section provides the relevant background required to develop our generalized waveguide theory. We begin with a brief review of Maxwell's equations and the concept of electromagnetic duality. We then show how duality naturally leads to the idea of the general mixed-monopole material
and discuss how metamaterials can be thought of as an approximate mixed-monopole material. Finally, we review waveguide theory for the slab and cylindrical geometries.

\subsection{\label{duality}Maxwell's Equations and Electromagnetic Duality}

In source-free regions, Maxwell's equations have a symmetric form, and the duality of Maxwell's equations becomes apparent. These equations can also be written in regions containing charges and currents, but the presence of electric charges and absence of magnetic ones leads to an asymmetry in Maxwell's equations~\cite{Jackson1999}. Restoring the symmetry of Maxwell's equations can be done by mathematically allowing for the existence of magnetic monopoles~\cite{Rajantie2012, LeBlanc2014}. 

Considering both electric and magnetic charges (by inserting terms for a magnetic charge density and current density) allows Maxwell's equations to be written as follows~\cite{Jackson1999}:
\begin{equation}
\begin{split}
&\nabla\cdot \bm D =\rho_{\text{e}},\ \ \ \ \ \ \ \ \ \ \ \ \ \ \ \nabla\cdot \bm B=\rho_{\text{m}},\\
&\nabla\times \bm E=-\frac{\partial \bm B}{\partial t}+\bm J_{\text{m}},\ \nabla\times \bm H=\frac{\partial \bm D}{\partial t}+\bm J_{\text{e}}\label{eq:Maxwell},
\end{split}
\end{equation}
where $\bm E$ is the electric field, $\bm D$ is the electric displacement field, $\bm B$ is the magnetic induction field and $\bm H$ is the auxiliary magnetic field. The free magnetic and electric current densities are represented by $\bm J_{m}$ and $\bm J_{e}$, respectively, and $\rho_{m}$ and $\rho_{e}$ are free magnetic and electric charge densities, respectively~\cite{Jackson1999}. To ensure symmetry, we require the magnetic monopole to behave in a manner analogous to the electric charge~\cite{Rajantie2012}.

In this paper, we assume linear, homogeneous and isotropic materials to simplify the expressions, allowing us to focus on duality. The components of the auxiliary fields in these types of materials are~\cite{Jackson1999}:
\begin{equation}
\begin{split}
&\bm D(\omega)\stackrel{\text{def}}{=}\varepsilon(\omega)\bm E(\omega),\\
&\bm B(\omega)\stackrel{\text{def}}{=}\mu(\omega) \bm H(\omega)\label{eq:EB},
\end{split}
\end{equation}
with $\varepsilon(\omega)$ and $\mu(\omega)$ the frequency-dependent electric permittivity and magnetic permeability of the material. The permeability of most materials is constant, but for certain materials, such as magnetic materials, it can be frequency-dependent.

Equations~(\ref{eq:Maxwell}) and~(\ref{eq:EB}) reveal that for systems in which the symmetric Maxwell's equations are valid, electric parameters, including field, current density, charge or any parameter related to them, can be easily exchanged to magnetic parameters, without changing the form of Maxwell's equations, and vice versa.
This EM duality can be manifest in any structure in which symmetric Maxwell's equations exist. If a material were to feature  electromagnetic duality, we would expect to find within it both electric and magnetic monopoles, as can be seen from Eq.~(\ref{eq:Maxwell}). 

\subsection{\label{Materials}{Mixed-Monopole Materials}}

Mixed-monopole materials, materials that contain both electric and magnetic monopoles, can expand the parameter space for electromagnetic phenomena such as negative, positive and zero permeability and permittivity.  Although magnetic monopoles are not known to exist, they are theoretically permitted and serve as a valuable mathematical tool~\cite{Dirac1931}. The electric and magnetic properties of materials are encompassed in the permittivity and permeability of the medium, respectively.

Different models are presented in the literature to characterize frequency-dependent permittivity and permeability.
However,models should exhibit all the EM features of materials, with refractive index ranging from double-positive to double-negative values. One model that can be used for~$\varepsilon$ and~$\mu$ in mixed-monopole materials is the Drude-Lorentz model~\cite{Jackson1999}. This model is a good candidate for describing EM susceptibilities of mixed-monopole materials, as it is derived from a microscopic description of the response of a material to an incident electromagnetic field~\cite{Jackson1999}. 

Electric permittivity is the result of a material polarizing in response to an external electric field. The Drude-Lorentz form for the frequency-dependent permittivity of a mixed-monopole material is~\cite{Jackson1999}
\begin{equation}
	\frac{\varepsilon(\omega)}{\varepsilon_0}
		=1+\frac{F_{\textrm{e}} \omega_{\textrm{e}}^2}{\omega_{\textrm{0}_{\textrm{e}}}^2
			-\omega^2+\text{i}\Gamma_{\textrm{e}}\omega}
\label{eq:epsilondef},
\end{equation}
where $F_{\textrm{e}}$ is electric oscillation strength, $\omega_{\textrm{0}_{\textrm{e}}}$ is electric resonance frequency, $\Gamma_{\textrm{e}}$ is the electric damping constant, and all describe charge-Ion dipole oscillation
(Drude-Lorentz model). The angular frequency of the incident electromagnetic field is $\omega$.
In this equation $\omega_{\textrm{e}}$ is the electric plasma frequency and is
\begin{equation}
	\omega_{\textrm{e}}
		\stackrel{\text{def}}{=}\sqrt{\frac{N_{\textrm{e}} q_{\textrm{e}}^2}{m_{\textrm{e}} \varepsilon_0}}, \label{eq:electricplasma}
\end{equation}
where $N_{\textrm{e}}$ is the number density of electrons, $q_{\textrm{e}}$ and $m_{\textrm{e}}$ are electron charge and mass and $\varepsilon_0$ is vacuum permittivity~\cite{Jackson1999}.

The magnetic susceptibility of a mixed-monopole material
is, also, described by a frequency-dependent function, which is called the permeability. The frequency-dependent permeability is generated by magnetic dipole oscillations in mixed-monopole materials. 
As we require magnetic monopole behavior to mirror that of electric charges, the permeability, like the permittivity, is described by the Drude-Lorentz model:
\begin{equation}
	\frac{\mu(\omega)}{\mu_0}
		=1+\frac{F_{\textrm{m}} \omega_{\textrm{m}}^2}{\omega_{\textrm{0}_{\textrm{m}}}^2-\omega^2+\text{i}\Gamma_{\textrm{m}}\omega}
\label{eq:mudef}
\end{equation}
with $\omega_{\textrm{0}_{\textrm{m}}}$ the magnetic resonance frequency~\cite{Lavoie2012, Penciu2010, Pendry1999}, $F_{\textrm{m}}$ is the magnetic oscillation strength and $\Gamma_{\textrm{m}}$ the magnetic damping constant, which are all parameters of magnetic monopole-monopole coupling interaction in mixed-monopole materials.
In Eq.~(\ref{eq:mudef}) the magnetic plasma frequency~$\omega_{\textrm{m}}$ obeys
\begin{equation}
	\omega_{\textrm{m}}
		\stackrel{\text{def}}{=}
			\sqrt{\frac{N_{\textrm{m}} q_{\textrm{m}}^2}{m_{\text m} \mu_0}}
\label{eq:magneticplasma}
\end{equation}
with $N_{\textrm{m}}$ the number density of magnetic charges, $q_{\textrm{m}}$ and $m_{\textrm{m}}$ the magnetic charge and mass, and $\mu_0$ the vacuum permeability. The complex-valued permittivity and permeability are denoted as $\varepsilon=\varepsilon'+\text{i}\varepsilon''$ and $\mu=\mu'+\text{i}\mu''$, respectively.

The two expressions~(\ref{eq:epsilondef}) and~(\ref{eq:mudef}) are identical up to converting between electric and magnetic symbols.
This similarity motivates us to introduce one symbol for both electric permittivity and magnetic permeability:
\begin{equation}
\label{eq:zetaepsilonmu}
	\zeta_{\text e}
		\stackrel{\text{def}}{=}\varepsilon,\;
	\zeta_{\text m}
		\stackrel{\text{def}}{=}\mu.
\end{equation}
We can use this convenient notation to refer to either~$\varepsilon$ or~$\mu$
as~$\zeta$, with the meaning made clear by context.
Accordingly, we can use the parameters
\begin{equation}
\label{eq:FomegaGamma}
	F_\text{e,m},\;
	\omega_\text{e,m},\;
	\omega_{\textrm{0}_{\textrm{e,m}}},\;
	\Gamma_\text{e,m}
\end{equation}
as the electric and magnetic components of oscillation strength, plasma frequency, resonance frequency and damping, respectively.

The complex refractive index of the lossy mixed-monopole material is
\begin{equation}
	n
		\stackrel{\text{def}}{=}
			\sqrt{\frac{\varepsilon \mu}{\varepsilon_0\mu_0}}=n_{\textrm{r}}+\text{i}n_{\textrm{i}}
\label{eq:n}.
\end{equation}
As square roots have two possible solutions, one positive and one negative, we must choose the appropriate case. We are assuming passive materials and so choose the sign that corresponds to $n_{\textrm{i}}>0$ to ensure that the material does not have gain. This choice leads to $n_{\textrm{r}}<0$ at frequencies that have both $\varepsilon'<0$ and $\mu'<0$~\cite{Lavoie2012}. 

Although materials containing both magnetic and electric monopoles are not known to exist in nature, we can approximate the desired behavior with a material that behaves the same way on a large enough scale.  
For instance, artificially engineered metamaterials, which are examples of mixed-monopole materials and exhibit both electric and magnetic effects, can be used to simulate the presence of magnetic monopoles~\cite{Penciu2010}. As metamaterials approximate the response of mixed-monopole materials, their EM properties can be studied using the Drude-Lorentz model. 

\subsection{\label{Metamaterials}{Metamaterials}}

Metamaterials are artificially engineered materials~\cite{ Shalaev2007} that have properties and functionalities previously not attainable in natural materials~\cite{Ves68, Smith2004, Pendry2000, Padilla2006}. 
Recently, metamaterials with different structures and properties are constructed and investigated~\cite{Ves68, Boltasseva2008, Pendry1999, Smith2000, Burgos2010}. Developments in metamaterial science have provided the possibility to realize materials with positive, zero and negative effective refractive indices~\cite{Smith2000, Vesseur2013, Islam2015, Schilling2011, Maas2013}. Zero-refractive-index metamaterials have infinite wavelength, quasi-infinite phase velocity and also quasi-uniform phase of light in the material. The quasi-uniform phase of light implies that the dipoles in the metamaterial are oscillating uniformly~\cite{Schilling2011}.

Now we consider metamaterials constructed solely with metal wires, meshes or plates~\cite{Penciu2010}.
For such metamaterials,
bulk permittivity is the dominant electric response. 
Therefore, the electric resonance frequency~$\omega_{0_\text{e}}$
is zero so the Drude-Lorentz model~(\ref{eq:epsilondef}) reduces to 
\begin{equation}
	\frac{\varepsilon(\omega)}{\varepsilon_0}
		=1-\frac{\omega_{\textrm{e}}^2}{\omega(\omega
			+\text{i}\Gamma_{\textrm{e}})},\label{eq:epsilondefmeta}
\end{equation}
which is the Drude model for electric susceptibility~\cite{Jackson1999, Lavoie2012}.

Whereas $\omega_{0_\text{e}}=0$ for the electric susceptibility,
the magnetic resonance frequency~$\omega_{0_\text{m}}$ is not necessarily zero in Eq.~(\ref{eq:mudef}).
The nonzero value of~$\omega_{0_\text{m}}$ is due to the sub-wavelength structure of the metamaterial components
responsible for its magnetic susceptibility.
For example,
the fishnet metamaterial~\cite{Penciu2010, Smith2000},
which we analyze here,
the magnetic response is due to current loops in both the slabs and the necks of the structure.

As metamaterials derive their magnetic response from their structure, rather than from magnetic monopoles, the parameters in Eq.~(\ref{eq:mudef}) are not functions of monopole-monopole interactions. Most notably, the plasma frequency in a metamaterial is simply the frequency of the incident field,
$\omega_\text{m}=\omega$. The monopoles in a material will naturally oscillate at the plasma frequency when perturbed. A metamaterial does not have monopoles, but magnetic dipoles created by current loops. As these current loops are driven by the incident field, they will tend to oscillate at the driving frequency, i.e.\ that of the incident field. The remaining parameters in Eq.~(\ref{eq:mudef}) reflect the effects of the metamaterial geometry, as well as the resistance of the metal, and the capacitance and inductance created between the plates~\cite{Penciu2010}.

\subsection{\label{SCwaveguides}Waveguides}

Waveguides are structures that have the ability to confine and direct the energy of EM waves. Waveguides can be made of different materials and can have different structures, such as rectangular, slab and cylindrical geometries. The characteristics and behavior of waveguide modes depend on both the geometry and materials~\cite{Yeh08}.

\begin{figure}
\begin{tikzpicture}
\begin{scope}[shift = {(4.6,-.2)}]
  \node[cylinder,draw=black,aspect=8.6,  rotate=142, 
        minimum height=4.4cm,minimum width=2cm,
        shape border rotate=90,
        cylinder uses custom fill,
        cylinder body fill=gray!50,
        cylinder end  fill=gray!50]
   () {};
  \draw[fill=white] (-1.21,-1.56) circle[x radius=.6,y radius=.6];
  \draw [->,thick] (-.5,-.5) -- (.5,.7);
  
  \draw [->,thick] (-1.3,-1.6) -- (-.86,-1.1);
  \node [below] at (-.95,-1.25) {\large $a$};
  
  \node [below] at (-5,-2.7) {\large $(a)$};
  \node [below] at (-1.2,-2.7) {\large $(b)$};
  
  \node [below] at (.3,0.2) {\Large $\bold{\it{z}}$};
  
  \node [below] at (-1.6,-1.3) {\large $1$};
   \node [below] at (-2,-1.3) {\large $2$};
  \end{scope}
  
 \begin{scope}[shift = {(-3,-.2)}]
 
\pgfmathsetmacro{\cubex}{1.7}
\pgfmathsetmacro{\cubey}{.65}
\pgfmathsetmacro{\cubez}{4.5}

\draw[black,fill=gray!50] (3.6,-1.7,0) -- ++(-\cubex,0,0) -- ++(0,-\cubey,0) -- ++(\cubex,0,0) -- cycle;
\draw[black,fill=gray!50] (3.6,-1.7,0) -- ++(0,0,-\cubez) -- ++(0,-\cubey,0) -- ++(0,0,\cubez) -- cycle;
\draw[black,fill=gray!50] (3.6,-1.7,0) -- ++(-\cubex,0,0) -- ++(0,0,-\cubez) -- ++(\cubex,0,0) -- cycle;

\draw[black,fill=gray!50] (3.6,-0.5,0) -- ++(-\cubex,0,0) -- ++(0,-\cubey,0) -- ++(\cubex,0,0) -- cycle;
\draw[black,fill=gray!50] (3.6,-0.5,0) -- ++(0,0,-\cubez) -- ++(0,-\cubey,0) -- ++(0,0,\cubez) -- cycle;
\draw[black,fill=gray!50] (3.6,-0.5,0) -- ++(-\cubex,0,0) -- ++(0,0,-\cubez) -- ++(\cubex,0,0) -- cycle;

    \node [below] at (3.4,-.6) {\large $2$};
   \node [below] at (3.4,-1.18) {\large $1$};
   \node [below] at (3.4,-1.8) {\large $2$};
   
   \draw [<->,thick] (1.8,-1.68) -- (1.8,-1.2);
   \node [below] at (1.55,-1.2) {\large $w$};
   
\end{scope}
   \end{tikzpicture}
   \caption{\label{fig:slabandcylinguide} Diagram of the slab and cylindrical waveguides. Region 1 in both guides is for the core made of lossless dielectric and region 2 is cladding based on an mixed-monopole material, metamaterials in this case. The width of the slab guide is $w$ and the radius of the cylindrical guide is $a=w/2$. The propagation direction of the fields is along the $z$ axis.}
\end{figure}
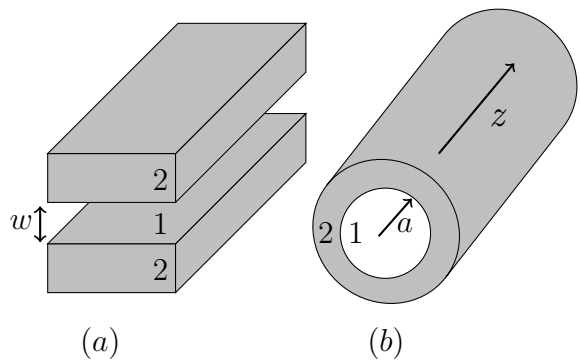

In order to study the EM properties of waveguides in this paper, slab and cylindrical geometries are considered, as examples. Waveguides having a slab geometry can support two different mode types, transverse electric (TE) and transverse magnetic (TM). Cylindrical guides support both TE and TM, as well as a third mode type, which is neither TE or TM, called a hybrid (HE) mode~\cite{Yeh08}. A schematic diagram of a slab and cylindrical waveguide is illustrated in Fig.~\ref{fig:slabandcylinguide}.

To characterize different modal behavior in the waveguides, the dispersion relation is needed, which is attained by solving Maxwell's equations in both the core and cladding, and then applying the appropriate boundary conditions for the geometry~\cite{Yeh08, Jackson1999}. The dispersion relations of TM and TE modes
for the symmetric slab guide are~\cite{Yeh08, Lavoie2012}
\begin{equation}
	\frac{2\zeta_1\zeta_2}{\gamma_1\gamma_2}
		=-\left(\frac{\zeta_1^2}{\gamma_1^2}+\frac{\zeta_2^2}{\gamma_2^2}\right)\tanh(\gamma_1w)
\label{eq:zetadispersion},
\end{equation}
with~$\zeta$ in Eq.~(\ref{eq:zetadispersion})
referring to both the electric permittivity and to the magnetic permeability~(\ref{eq:zetaepsilonmu}).
In Eq.~(\ref{eq:zetadispersion})
\begin{equation}
	\gamma_j
		=\sqrt{\tilde\beta^2-\omega^2\varepsilon_j\mu_j}
\end{equation}
is the complex wave number for the transverse components of the fields, $j=1, 2$
(1 and 2 indicate the core and cladding, respectively). Here, $\tilde\beta=\beta+\text{i}\alpha$ is the complex propagation constant (in the propagation direction) with $\beta$ and $\alpha$ the dispersion and attenuation, respectively,
and $w$ is the core width.

For the HE modes in a cylindrical guide, the dispersion relation is given by~\cite{Yeh08, Lavoie2012}
\begin{equation}
\begin{split}
&\left(\frac{1}{\kappa^2_1}+\frac{1}{\gamma^2_2}\right)^2= \left(\frac{\mu_1}{\kappa^2_1}\frac{J'_{m}(a\kappa_1)}{J_{m}(a\kappa_1)}+\frac{\mu_2}{\gamma^2_2}\frac{K'_{m}(a\gamma_2)}{K_{m}(a\gamma_2)}\right)\\
&\times\left(\frac{\varepsilon_1}{\kappa^2_1}\frac{J'_{m}(a\kappa_1)}{J_{m}(a\kappa_1)}+\frac{\varepsilon_2}{\gamma^2_2}\frac{K'_{m}(a\gamma_2)}{K_{m}(a\gamma_2)}\right) \left(\frac{a^2 \omega^2}{\tilde\beta^2 m^2}\right) \label{eq:cylindisp},
\end{split}
\end{equation}
with $J_{m}$ and $K_{m}$ the Bessel and modified Bessel functions, respectively, and the prime representing the derivative of $J_{m}$ and $K_{m}$ with respect to $r$ (radial coordinate). Here $\gamma_2$ and $\kappa_1=i \gamma_1$ are the wave numbers for the cladding and core, respectively, $a$ is the core radius and $m\geq0$ is an integer characterizing azimuthal symmetry. 

We can recover the dispersion relations for the TE and TM modes of a cylindrical waveguide by setting $m=0$ in Eq.~(\ref{eq:cylindisp}), which reduces it to two separate dispersion equations, one for TE and the other for TM modes. For the $m>0$ case, hybrid modes are needed to satisfy the boundary conditions~\cite{Yeh08}. As the TM and TE modes of the cylindrical guide are similar to those of the slab guide, we focus on the HE modes. These dispersion relations describe the modal dispersion and can be solved to obtain the complex propagation constant. 

As metamaterials are examples of mixed-monopole materials, metamaterial waveguides can be studied as examples for more general waveguides based on mixed-monopole materials. Due to wide applicability of optical waveguides, metamaterial waveguides have been recently studied in order to extend waveguides' range of application even further.
The goal for metamaterial waveguides is to realize optical properties significantly beyond conventional dielectric and metal waveguides~\cite{Lavoie2012, Kim2007, DAguanno2005, Shadrivov2003}.

In order to exploit fully the promise of metamaterial waveguides,
we need to understand their nature and, in particular, the types of modes that are supported.
Metamaterial waveguides can support three distinct mode behaviors:
ordinary, surface plasmon polariton (SPP) and SPP-ordinary hybrid modes that are the combination of SPP and ordinary modes. Ordinary modes are related to travelling EM fields inside the core, whereas for SPP modes the energy is concentrated around the core-cladding interface~\cite{Lavoie2012}.
To decrease the complexity of our results, we focus on ordinary modes in this paper. 

\section{\label{Approach}Approach}

We generalize the Drude-Lorentz model used for the resultant permittivity and permeability 
in order to explore duality and its effects on the waveguide at the microscopic level.
Although we recognize that the waveguides can comprise both material and metamaterial components,
our analysis at the microscopic level treats the magnetic properties as arising from magnetic monopoles
in the medium, analogous to plasmonic effects arising from electric monopoles (namely, electrons) in the medium.
Our treatment of electric and magnetic monopoles is discuss in Subsec.~\ref{subsec:material},
and we develop the generalized Drude-Lorentz model in Subsec.~\ref{subsec:generalized}.

Our incorporation of magnetic monopoles into the theory is not meant to imply that we are aiming to use magnetic monopoles in construction of waveguides but rather to exploit a manifestation of electromagnetic duality at the microscopic level by treating the medium as having magnetic monopoles.
On the macroscopic level, this magnetic monopole description of permittivity and permeability is valid as an effective theory.

In Subsec.~\ref{subsec:numericalmethod}, we present our numerical method to solve the dispersion equations of slab and cylindrical waveguides considerd in this paper. By numerically solving the dispersion equations and obtaining the complex propagation constant of waveguides, we investigate the EM characteristics of slab and cylindrical waveguides (Fig.~\ref{fig:slabandcylinguide}) constructed from mixed-monopole materials using the parameters introduced in Subsec.~\ref{subsec:characterization}. 

\subsection{\label{subsec:material}Mixed-Monopole Materials}

In this subsection we explain the nature of mixed-monopole materials,
which include both electric and magnetic monopoles.
In this subsection,
we introduce and discuss the Drude-Lorentz model
and describe the electromagnetic response of mixed-monopole materials.
Then we introduce certain materials as representative examples of mixed-monopole materials.

We consider mixed-monopole materials in this paper, which comprise electric and magnetic monopoles with positive and negative charges. Rather than refer to charges, we generalize to the terms electric monopole and magnetic monopole for the negative charges, and we refer to the massive positive charges as electric and magnetic ions. Our treatment can be extended to dipoles with low-mass negative charges ions and high-mass positive charges. We recognize that in metamaterials the magnetic dipoles are actually generated by electric currents. Nevertheless the notion of valence magnetic monopoles and oppositely charged massive magnetic nuclei is convenient to build a fully dualistic description.

We employ the Drude-Lorentz model for both electric and magnetic susceptibilities~\cite{Penciu2010}. This model can describe both dielectric and metallic materials. In the limit of weak coupling between charges and ions, the Drude-Lorentz model reduces to the Drude model for metals~\cite{Engheta2006}: negative charges are detached from the ions, allowing them to move more freely within the material. The Drude and Drude-Lorentz models assume that inter-electron interactions are negligible~\cite{Jackson1999}.

We extend this conventional electrical approach to incorporate
magnetic monopole-monopole interactions. This is justified by the fact that the Drude-Lorentz model assumes minimal inter-dipole interaction (i.e.\ there is enough separation between the atoms that nearest-neighbour interaction is negligible compared to the effects of the incident field);
thus, we also assume a large inter-monopole separation consistent with the large inter-monopole separation for the electric case.

In this extended Drude-Lorentz model, various parameters are included such as the mass of the monopoles.
The effective masses of the electric and magnetic monopoles depend, in part, on the band structure created by the lattices formed from the massive electric and magnetic ions. Determining the band structure and effective masses would require a quantum treatment, but is not germane to our work.

In this subsection we have shown the properties of mixed-monopole materials and the capability of Drude-Lorentz model for describing them. With this knowledge, we are now equipped to generalize the Drude-Lorentz model applicable to linear, homogeneous isotropic materials for application to mixed-monopole materials.

\subsection{\label{subsec:generalized}Generalized Drude-Lorentz model}

In this subsection, we construct the Drude-Lorentz model for a mixed-monopole material.
Then we show how this model can be used to determine the permittivity and permeability of the medium and furthermore show how electromagnetic duality manifests in these expressions.
We use the relations for the generalized Drude-Lorentz model and refractive index to show different kinds of materials with positive, negative, near-zero and zero refractive index.

We introduce a modified form of the Drude-Lorentz model, called the generalized Drude-Lorentz model, to facilitate studying duality in waveguides. The generalized model allows us to investigate EM symmetries of systems under the exchange of material properties from electric to magnetic and vice versa. The generalized Drude-Lorentz model has the form
\begin{equation}
	\frac{\zeta_\text{e,m}}{\zeta_{0_\text{e,m}}}=\zeta_{\textrm{b}_\text{e,m}}+\frac{F_\text{e,m} \omega_\text{e,m}^2}{\omega_{0_\text{e,m}}^2-\omega^2+\text{i}\Gamma_\text{e,m} \omega}\label{eq:dualmodel},
\end{equation}
where $\zeta_\text{e,m}$ is permittivity or permeability, $\zeta_{0_\text{e,m}}$ is vacuum permittivity or permeability, $\zeta_{\textrm{b}_\text{e,m}}$ is the background permittivity or permeability, 
and the other parameters~(\ref{eq:FomegaGamma}) have been explained earlier.
Subscripts ${}_\textrm{e}$ and ${}_\textrm{m}$ refer to electric and magnetic parameters, respectively. 

Our generalized model respects the duality of electromagnetism and describes the plasmonic effects of both electric and magnetic monopoles. We see this mathematical formalism as being inspired by treating formally a magnetic monopole plasma. To study EM duality in the system, we replace~$\varepsilon$~(\ref{eq:epsilondef})
and~$\mu$~(\ref{eq:mudef}) 
by~$\zeta_\text{e,m}$, respectively, in the generalized-model dispersion relations.

By applying duality transformations (exchanging EM parameters), we can realize the duality of electromagnetism in different systems. The Drude-Lorentz model (and possibly others) can be recovered from the generalized Drude-Lorentz model through appropriate parameter choices. This model has the ability to describe materials with all possible values for refractive index.

The refractive index can be expressed in terms of the generalized model as
\begin{equation}
n=\sqrt{\frac{\zeta_{\textrm{e}}\zeta_{\textrm{m}}}{\zeta_{0 \textrm{e}}\zeta_{0 \textrm{m}}}}\label{eq:refractive}.
\end{equation}
This equation, along with Eq.~(\ref{eq:dualmodel}), allows for positive, negative, near-zero and zero-refractive-index materials, given the appropriate parameter values. The condition for obtaining a refractive index of zero depends on where $\zeta_{\textrm{e}}$ and $\zeta_{\textrm{m}}$ become zero, so there are generally four frequencies at which zero-refractive-index occurs. There are also regions with near-zero refractive index.

In summary, expressions for generalized Drude-Lorentz model~(\ref{eq:dualmodel}) and for refractive index~(\ref{eq:refractive}) manifest the duality of electromagnetism. The generalized Drude-Lorentz model can show exotic material properties with even negative and zero refractive indices.

\subsection{\label{subsec:characterization} Characterization}

In this subsection, we introduce dispersion equations and suitable waveguide parameters that augment the material parameters introduced so far.
Then we proceed to construct equations in order to determine the values of these parameters.
These values are obtained by solving the dispersion equations in two cases,
namely, for slab and cylindrical waveguides as they are easily solved examples and provide a clear foundation for studying more complicated cases.

We can rewrite dispersion relations for TE and TM modes in slab guide using the generalized model, as Eq.~(\ref{eq:zetadispersion}), where we suppressed the subscript on $\zeta_\text{e,m}$ for readability. In  Eq.~(\ref{eq:zetadispersion}) $\zeta_j$ becomes $\varepsilon_j$ for TM modes and $\mu_j$ for TE modes. As $\gamma_j$ contains only the product of~$\varepsilon$ and~$\mu$, it is invariant under the duality transformation. This unified equation makes the duality between TE and TM modes evident. Therefore, the generalized model provides a unified relation that can be used to calculate both mode types. We see from Eq.~(\ref{eq:zetadispersion}) that the dispersion relation for the TE (TM) modes of the slab guide can be transformed into the dispersion relation for the TM (TE) modes by replacing the magnetic version of $\zeta$ with the electric one and vice versa.

For HE modes the unified dispersion equation is:
\begin{equation}
\begin{split}
&\left(\frac{1}{\kappa^2_1}+\frac{1}{\gamma^2_2}\right)^2=
\left(\frac{\zeta_{\textrm{m} 1}}{\kappa^2_1}\frac{J'_{m}(a\kappa_1)}{J_{m}(a\kappa_1)}+\frac{\zeta_{\textrm{m} 2}}{\gamma^2_2}\frac{K'_{m}(a\gamma_2)}{K_{m}(a\gamma_2)}\right)\\
&\times\left(\frac{\zeta_{\textrm{e} 1}}{\kappa^2_1}\frac{J'_{m}(a\kappa_1)}{J_{m}(a\kappa_1)}+\frac{\zeta_{\textrm{e} 2}}{\gamma^2_2}\frac{K'_{m}(a\gamma_2)}{K_{m}(a\gamma_2)}\right)\left(\frac{a^2 \omega^2}{\tilde\beta^2 m^2}\right)\label{eq:cylindispnew},
\end{split}
\end{equation}
where $\kappa_1$ is the complex wave number in the core ($\kappa_1=i \gamma_1$) and $\gamma_2$ is the complex wave number for the cladding (subscripts 1 and 2 are for core and cladding regions of the cylindrical guide, respectively). Just like the slab guide, the dispersion relation for TE (TM) modes can be transformed into that for the TM (TE) modes with a duality transformation. The dispersion relation for the HE modes, however, is left unchanged.

We can now characterize the waveguide parameters for various modes of the slab and cylindrical waveguides. 
The two waveguide parameters are relevant.
One parameter is the effective refractive index
\begin{equation}
\label{eq:neff}
	n_{\textrm{eff}}=\beta/ k_0,\;
	k_0=\omega/c =\lambda_0/2\pi
\end{equation}
of a guided mode,
which we see is directly proportional to the real part~$\beta$ of the propagation constant~\cite{Lavoie2012}:
The effective refractive index~(\ref{eq:neff}) is frequency dependent
and accounts for total modal dispersion
(i.e., combined effects of both chromatic and waveguide dispersion)
with the variations leading to different mode characteristics. 

We also need to consider mode attenuation in a waveguide, which accounts for intensity reduction in the direction of travel along the guide. Attenuation is mathematically described by the imaginary part of the complex propagation constant, which we denote as $\alpha$. As with the effective refractive index, the mode attenuation is frequency dependent
and is a combination of both material losses and energy leaking from the waveguide. These two parameters allow us to characterize the waveguide modes and how they are affected by duality.
We consider waveguides with a metamaterial or metal cladding and an air core. We then exploit the symmetry of the generalized model to study EM duality in these structures.

In this subsection, we have introduced two relevant waveguide parameters.
The two important parameters are effective refractive index of waveguide and waveguide attenuation
that are related to waveguides' propagation constants. 
Studying these parameters enables us to see waveguides behavior beside investigating EM duality in guides.  
However, the propagation constant is attainable by solving the transcendental dispersion equations of waveguides and requires numerical approaches for solving the equation. In the next subsection we briefly explain our numerical method for solving dispersion equations.  

\subsection{\label{subsec:numericalmethod}Numerical method}

As our waveguides' dispersion equations~(\ref{eq:zetadispersion}) and~(\ref{eq:cylindisp}) are transcendental,
they cannot be solved exactly and numerical techniques are required to solve the dispersion relations.
Various numerical approaches are available,
and we choose Newton's method
because exact expressions for derivatives are available and the technique converges.

Obtaining the complex propagation constant for slab and cylindrical waveguides requires the dispersion equations to be solved. As they are transcendental, we solve them numerically using the numerical root-finding function
Mathematica$^\circledR$'s {\tt FindRoot} function,
which is based on Newton's method.
Using this function, we find the complex propagation constant for the supported modes at different frequencies.

As multiple modes can propagate in a waveguide at a given frequency, different initial seed values are needed to obtain the propagation constant for different modes. Thus, we need to provide an initial seed value for each mode at a frequency where the value of $\tilde\beta$ is known approximately. We can then increment the frequency value and solve for $\tilde\beta$ by using the previously obtained value as a seed value for {\tt FindRoot}.
This procedure minimizes the possibility of obtaining a $\tilde\beta$ value for a different mode than the one for which we are solving and is repeated for all desired modes.

In this subsection, we explained the numerical method we use to solve the transcendental dispersion equation of slab and cylindrical waveguides to find the complex propagation constant for different modes in waveguides. By finding the propagation constant of waveguides, we can now investigate the different parameters of waveguides.

\subsection{Summary of Approach}

In this section we presented the properties of mixed-monopole materials and introduced the Drude-Lorentz model as an appropriate model to describe these materials. We then generalized the Drude-Lorentz model to a model that manifests the duality of electromagnetism and describes materials with refractive indices ranging from positive to negative values. 

Next we presented the dispersion equations of waveguides. The effective refractive index and attenuation are two relevant parameters that describe the waveguides and are attainable by numerically solving dispersion equations of waveguides.  Investigating these parameters enable the investigation of EM duality in waveguides. We finally presented our numerical method for solving the dispersion equations of slab and cylindrical waveguides.

\section{\label{Results} Results}

In this section, we present our results for bulk metamaterial,
which is described well by the behavior of mixed-monopole materials,
and for metamaterial waveguides.
This section's results yield clear physical insight regarding waveguide behavior
including metal and non-metal waveguide media.
We ensure that our results are consistent with existing theory.
To verify that the generalized model is correct and reliable, we recreated some previously obtained numerical results using the appropriate parameters~\cite{Lavoie2012}. To do so, we solved the modal dispersion relation of the symmetric slab and cylindrical guides and numerically determined the complex propagation constants. We compared our calculations and numerical results against previous work~\cite{Lavoie2012} using identical parameters.
Our results reproduce previous facts and satisfy the expected duality relations.

We first examine the effects of the duality transformation on the electromagnetic properties of a bulk metamaterial. We then apply the results to metamaterial-dielectric waveguides, to determine the behavior of select modes under the duality transformation. 

\subsection{\label{Metamaterials} Metamaterials}

\begin{figure}
\centering
\includegraphics[width=\columnwidth]{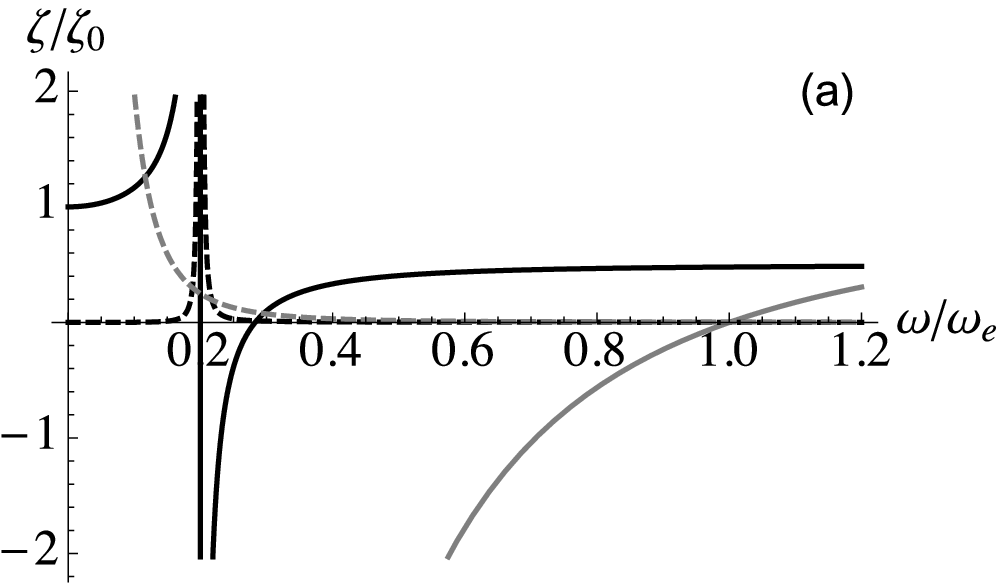}
\includegraphics[width=\columnwidth]{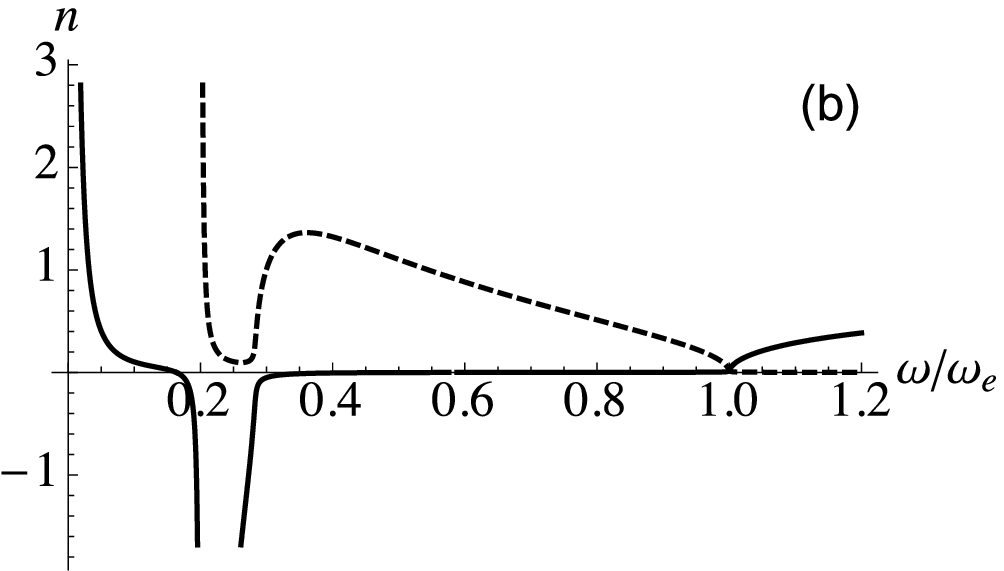}
\caption{\label{fig:epsilonandmu}Plots of (a)~$\zeta/\zeta_{0}$ that reduce to electric and magnetic responses: $\varepsilon/\varepsilon_0$ (black line) and $\mu / \mu_0$ (grey line), respectively, and (b) Refractive index of metamaterial as a function of frequency, using the generalized model. Solid and dashed lines represent real and imaginary parts, respectively.}
\end{figure}

To characterize the EM properties of metamaterials
we study the behavior of the permittivity, permeability and refractive index of a metamaterial
and depict numerical results inFig.~\ref{fig:epsilonandmu}.
These results are obtained by solving Eqs.~(\ref{eq:dualmodel}) and~(\ref{eq:refractive}).
The metamaterial parameters used are $\zeta_{\textrm{b}_\text{e,m}}:=1$, $F_{\textrm{e}}=1$, $F_{\textrm{m}}=0.5$, $\omega_{0_{\textrm{e}}}:=0$, $\omega_{0_{\textrm{m}}}:=0.2 \omega_{\textrm{e}}$, $\omega_{\textrm{e}}=1.37\times10^{16}~\text{s}^{-1}$, $\omega_{\textrm{m}}=\omega$ and $\Gamma_{\textrm{m}}=\Gamma_{\textrm{e}}=2.73\times10^{13}~\text{s}^{-1}$~\cite{Lavoie2012}.
These parameters represent a metamaterial with a metal-like permittivity, due to the dominant response of the metal used to construct the metamaterial, and a permeability that is induced through oscillating magnetic dipoles.

Figure~\ref{fig:epsilonandmu}(a) shows that the permittivity and permeability of metamaterials have positive, negative and zero values in different frequency regions. In the frequency regions where both~$\varepsilon$ and~$\mu$ are negative, the refractive index is also negative, which corresponds to a double-negative material
as shown in Fig.~\ref{fig:epsilonandmu}(b).
For our parameter choice, the material is double-negative in the frequency range
$0.2 \omega_{\textrm{e}}\lesssim \omega\lesssim 0.3\omega_{\textrm{e}}$. 

Therefore, from Figs.~\ref{fig:epsilonandmu}(a,b) we see that there are frequency regions in which the material has permittivity and permeability having positive, zero and negative values that let the refractive index have all possible values ranging from positive to negative values. Our model also solves $\varepsilon>0, \mu<0$ and double-zero refractive-index materials, however, because of the parameters choice in this paper we can not see these regions in the graphs.
Our generalized model thus describes materials with all possible values for refractive index.

By exchanging electric and magnetic properties in metamaterials, the behavior of $\epsilon$ and~$\mu$, in Fig.~\ref{fig:epsilonandmu}(a), is reversed, whereas the refractive index of the metamaterial remains unchanged. The reason for this behavior can be understood from the equation for the refractive index~(\ref{eq:n}), which contains only the product of~$\varepsilon$ and~$\mu$. Therefore, by exchanging electric and magnetic properties, which exchanges~$\varepsilon$ and~$\mu$, the behavior of refractive index does not change. These results show the EM duality of metamaterials by applying duality transformations.

\subsection{\label{Slabmetamaterial} Slab metamaterial-dielectric waveguide}

\begin{figure}
\centering
\includegraphics[width=\columnwidth]{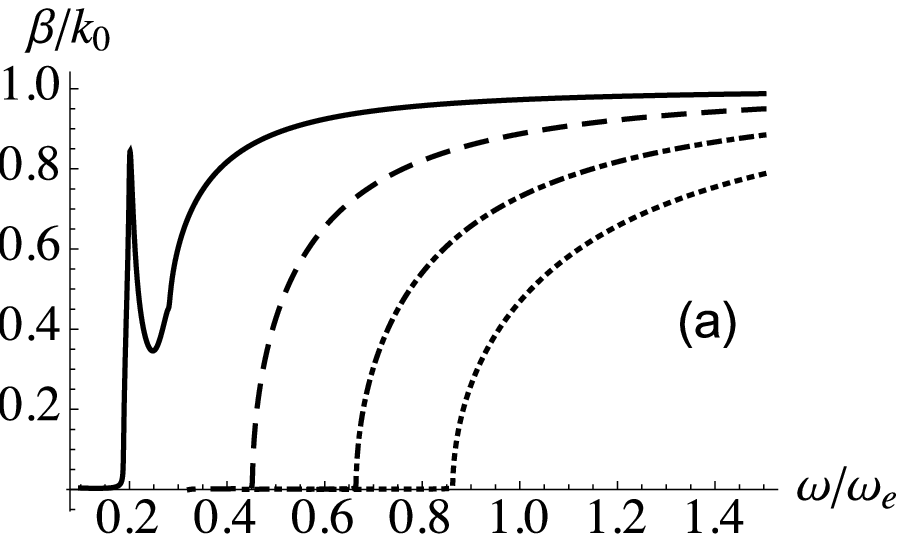}
\includegraphics[width=\columnwidth]{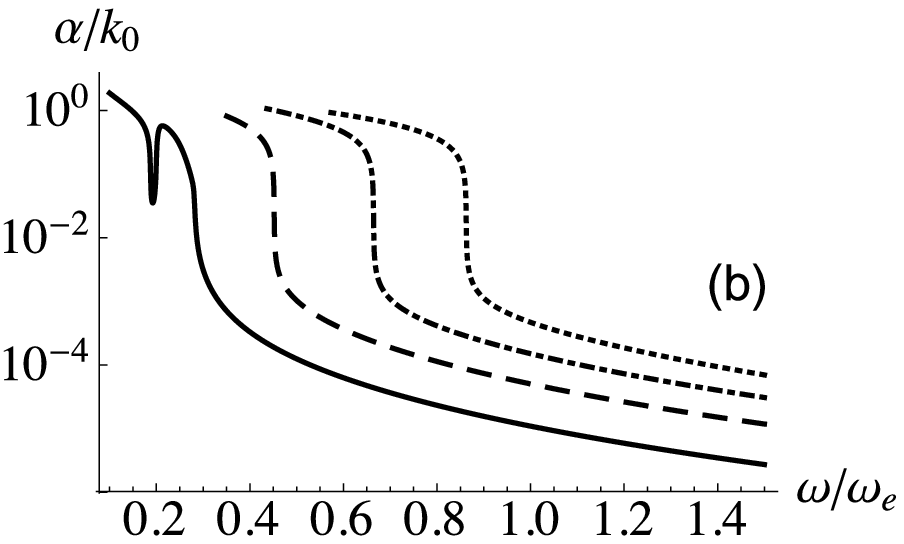}
\caption{\label{fig:SlabTEmode1} Plots of (a) effective refractive index, (b) attenuation, for different TE modes in the untransformed slab metamaterial-dielectric waveguide, using the generalized model. The solid, dashed, dot-dashed and dotted lines are for $\textrm{TE}_0$, $\textrm{TE}_1$, $\textrm{TE}_2$ and $\textrm{TE}_{3}$ modes, respectively.}
\end{figure}

The first waveguide example is a symmetric slab metamaterial-dielectric guide with hollow core and metamaterial cladding. We study the effective refractive index and attenuation of guided modes in this waveguide.
Figure~\ref{fig:SlabTEmode1} illustrates the behavior of the attenuation and $n_{\textrm{eff}}$ for the first few TE modes in the untransformed (before exchanging EM properties) slab metamaterial-dielectric waveguide as a function of frequency. 

\begin{figure}
\centering
\includegraphics[width=\columnwidth]{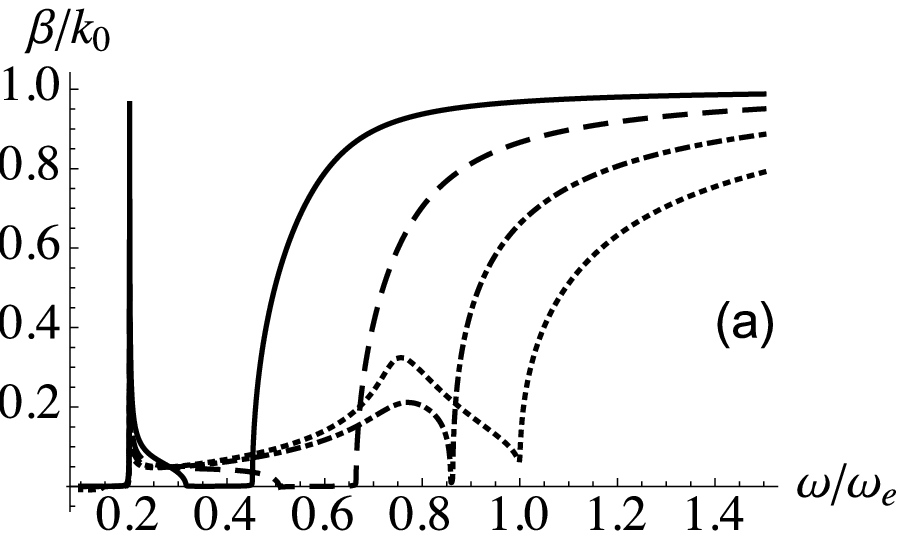}
\includegraphics[width=\columnwidth]{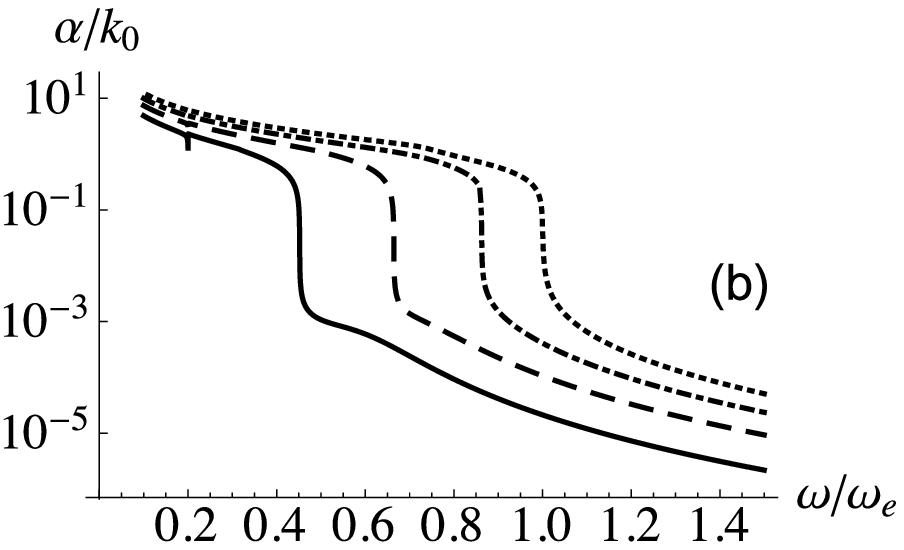}
\caption{\label{fig:SlabTMmode1} Plots of (a) effective refractive index, (b) attenuation for different $\textrm{TE}$ modes of the transformed slab metamaterial-dielectric waveguide, using the generalized model. The solid, dashed, dot-dashed and dotted lines are for $\textrm{TE}_0$, $\textrm{TE}_1$, $\textrm{TE}_2$ and $\textrm{TE}_{3}$ modes of transformed guide, respectively. 
}
\end{figure}

To test the EM symmetries of this slab waveguide, we exchange the permittivity and permeability on the materials. By applying this duality transformation in the slab metamaterial-dielectric waveguide, the behavior of the TE modes change in the transformed guide. Figure~\ref{fig:SlabTMmode1} shows the behavior of the TE modes in the transformed guide, which is identical to the TM modes in the untrasformed guide. This exchange of TE and TM modes under the duality transformation shows the EM symmetry of the slab metamaterial-dielectric waveguide. 

\subsection{\label{Cylindricalmetamaterial} Cylindrical metamaterial-dielectric waveguide}

\begin{figure}
\centering
\includegraphics[width=\columnwidth]{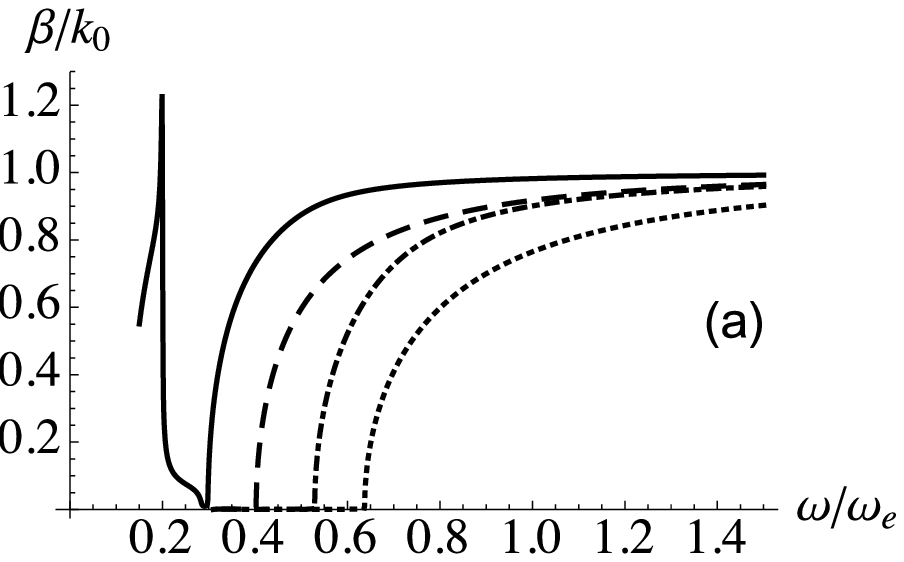}
\includegraphics[width=\columnwidth]{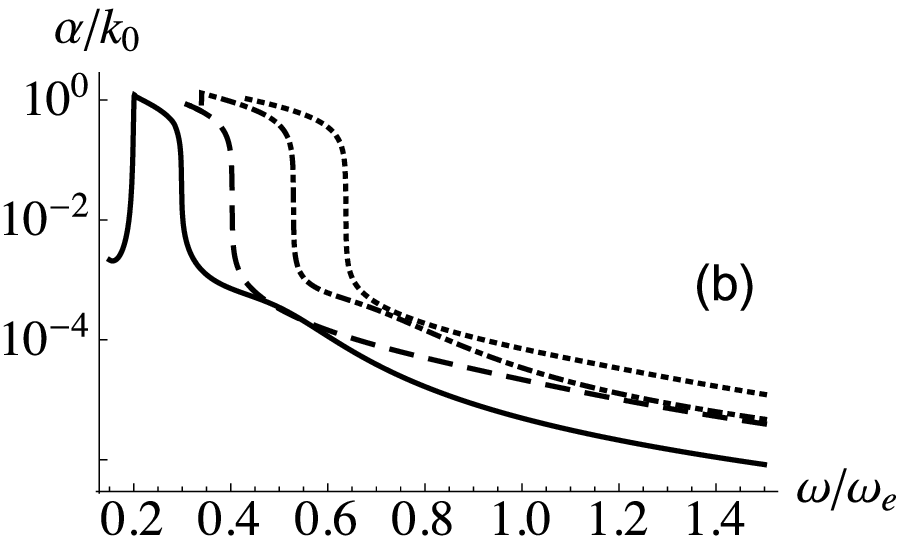}
\caption{\label{fig:cylinHEmode} Plots of (a) effective refractive index, (b) attenuation in both the untrasformed and trasformed cylindrical metamaterial-dielectric waveguide. The thick solid, dashed, dot-dashed and dotted lines are for $\textrm{HE}_{10}$, $\textrm{HE}_{11}$, $\textrm{HE}_{12}$ and $\textrm{HE}_{13}$ modes, respectively. The effective refractive index and attenuation remain unchanged under a duality transformation.}
\end{figure}

For the next example, we consider a cylindrical metamaterial-dielectric waveguide with a hollow core and metamaterial cladding. As with the slab guide, we investigate the effect of a duality transformation on select modes, HE in this case. The behavior of $n_{\textrm{eff}}$ and the attenuation of the modes is illustrated in Fig.~\ref{fig:cylinHEmode}.

These plots show that the HE mode behavior in a cylindrical metamaterial-dielectric guide shows a complete EM symmetry under a duality transformation. This is in contrast to the TE and TM modes, which exchange roles under a duality transformation, but expected given that the dispersion relation for these modes is invariant. However, the contributions of the electric and magnetic fields do exchange roles, as they depend on the permittivity and permeability in an asymmetric way. 

As for the slab guide, the dispersion relations for the TE and TM modes in the cylindrical guide convert to each other under the duality transformation. This means the TE and TM modes in the cylindrical guide have the same symmetric behavior as those in the slab guide when applying the duality transformation.  

\subsection{\label{Slabmetal} Slab metal-dielectric waveguide}

We have demonstrated the fact that metamaterials have a symmetric behavior under duality transformations. We have also shown that the slab and cylindrical metamaterial-dielectric waveguides respect EM duality. When we apply our generalized model to metal waveguides, we see consistent results with previous work~\cite{Lavoie2012}, showing the generalized model works for metals, as well as metamaterials. 

\begin{figure}
\centering
\includegraphics[width=\columnwidth]{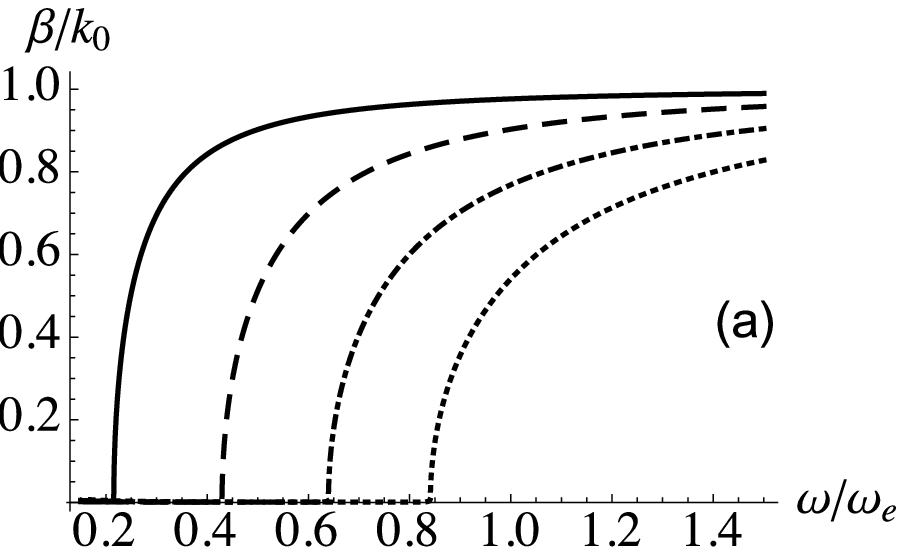}
\includegraphics[width=\columnwidth]{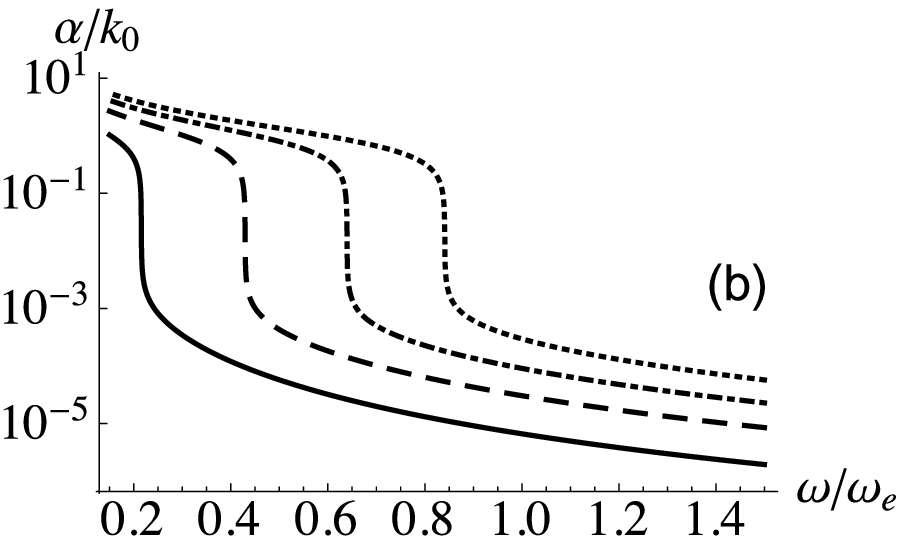}
\caption{\label{fig:SlabMetalGuideModes} Plots of (a) effective refractive index, (b) attenuation, for different TE modes in the untransformed slab metal-dielectric hollow waveguide, using the generalized model. The solid, dashed, dotted-dashed and dotted lines are for $\textrm{TE}_0$, $\textrm{TE}_1$, $\textrm{TE}_2$ and $\textrm{TE}_{3}$ modes, respectively.}
\end{figure}

We now perform the same procedure for the slab metal-dielectric waveguide, composed of a hollow core and metal cladding, to gain better insight on the generality of model. We first study the behavior of TE modes in the guide with the plots of effective refractive index and attenuation in Figs.~\ref{fig:SlabMetalGuideModes}(a,b).
We then perform the duality transformation on the waveguide. 

\begin{figure}
	\centering
	\includegraphics[width=\columnwidth]{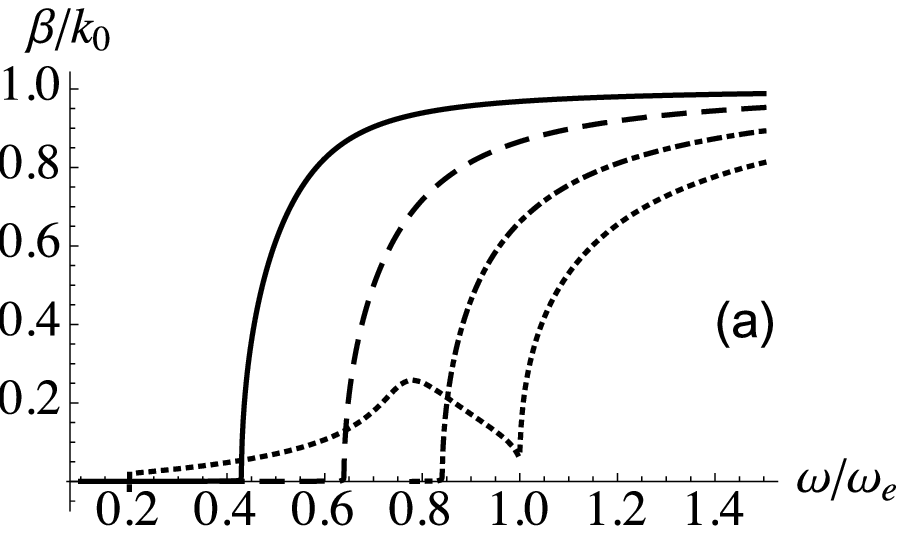}
	\includegraphics[width=\columnwidth]{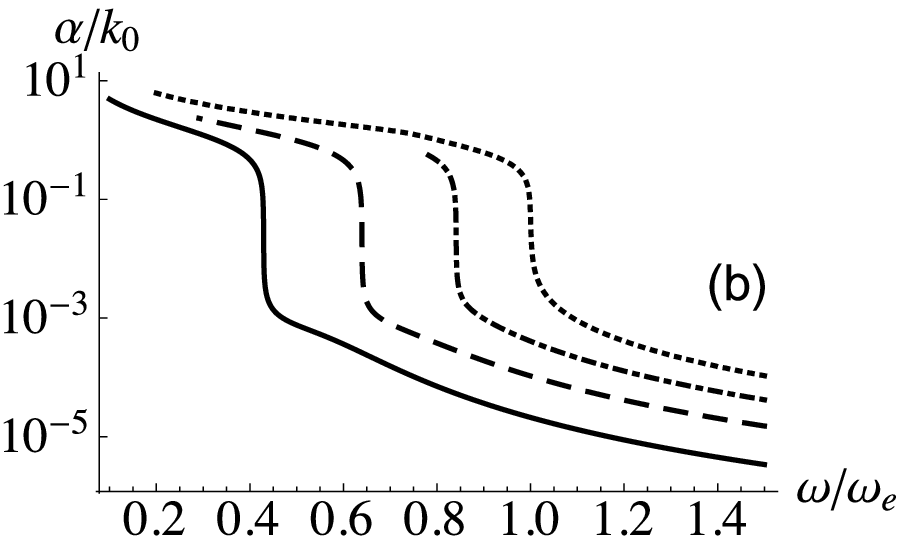}
	\caption{\label{fig:SlabMetalGuideModes2} Plots of (a) effective refractive index, (b) attenuation, for different TE modes in the transformed slab metal-dielectric hollow waveguide, using the generalized model. The solid, dashed, dotted-dashed and dotted lines are for $\textrm{TE}_0$, $\textrm{TE}_1$, $\textrm{TE}_2$ and $\textrm{TE}_{3}$ modes, respectively.}
\end{figure}

The physical interpretation of this transformation is that, by applying duality transformation, the metal cladding would have magnetic monopoles as free charge carriers, rather than electrons. This hypothetical ``magnetic metal'' would support the TM modes in transformed guide, which have the same behavior as TE modes in untransformed metal guide, as Figs.~\ref{fig:SlabMetalGuideModes2}(a,b).
These TE modes are the modes that the TM modes of the metal guide are transformed into under the duality transformation. For the HE modes in the metal-dielectric cylindrical guide, the duality transformation does not alter any of the plots, which means that this waveguide has EM symmetry under duality transformation.

\subsection{\label{summary} Summary}

In section IV, we investigated metamaterials, which are examples of mixed-monopole materials and are a generalization of standard materials to include both electric and magnetic responses.
Our results demonstrate the ability of our general model to provide a full description of materials, including double-positive, double-negative, positive-negative, negative-positive, permittivity-zero, permeability-zero and double-zero materials. 
We then investigated EM duality in waveguides constructed from metamaterials. Upon performing the duality transformation, the TE (TM) modes of a slab or cylindrical waveguide become TM (TE) modes, verifying the duality that is inherent in the generalized model. The HE modes in a cylindrical metamaterial-dielectric guide respect the EM duality by exchanging EM properties, though the effective refractive index and attenuation do not change. In other words, the TM modes of the untransformed guide are identical to the TE modes of the transformed guide, except that the E and B fields are exchanged, so we call them TE, rather than TM. The metal guides, as well as metamaterial waveguides, show EM symmetry under the duality transformations.

\section{\label{Conclusion}Conclusion}

In this paper, we develop a generalized theory for waveguides that respects the duality of electromagnetism. 
We demonstrate the usefulness of this theory by showing two concepts.
First, we show the ability of our generalized Drude-Lorentz model to describe general materials including mixed-monopole materials and metamaterials, along with waveguides based on these materials. 
Our generalized model underpins all physically accessible values of both permittivity and permeability,
and has the ability to include double-positive, positive and negative, double-negative and even zero-index materials. By exploiting the concept of magnetic monopoles, our theory can easily treat arbitrary combinations of general mixed-monopole materials, standard materials (such as metals and dielectrics) and metamaterials for selected waveguide geometries. 

Second, we show that if waveguides in one parameter regime are constructed, we can also identify the electromagnetically dual regime
with the same properties and make use of magnetic, rather than electric, properties. The study of EM symmetry in the systems is realizable by exchanging media from electric to magnetic material, by applying the duality transformations, and then studying the behavior of the refractive index of materials and the modes in the waveguides. This symmetric behavior is testable experimentally by investigating and comparing the EM properties of original and dual waveguides. 

Our theory gives us a good macroscopic description of the waveguide behavior and simplifies the calculations by 
using the duality of electromagnetism. We use the principle of duality in this paper to predict a waveguide that has the same behavior with the electric and magnetic fields swapped. In this case, we propose testing two waveguides with reversed parameters. 

Our results demonstrate that metamaterials exhibit a symmetric behavior under the duality transformation for all frequencies considered herein. This symmetry is due to the magnetic response of these materials, as well as their electric response, which simulates the existence of magnetic charges, and restores the duality of Maxwell's equations. Our model enables us to explore the EM symmetry of slab and cylindrical waveguide structures. This ability of our model can be used to investigate EM symmetries in other structures, as well. Our theory leads to a testable prediction for hollow-core metamaterial and metal waveguides that allows swapping the electric and magnetic properties.




\section{\label{sec:level1}Acknowledgment}

We acknowledge financial support from NSERC, AITF and the 1000 Talent Plan
of China.

\bibliography{Waveguideduality1}
\end{document}